\documentclass[12pt]{article}

\newcommand{\beq}{\begin{equation}}
\newcommand{\eeq}{\end{equation}}
\newcommand{\bea}{\begin{eqnarray}}
\newcommand{\eea}{\end{eqnarray}}
\newcommand{\third}{{\textstyle {1\over 3}}}
\newcommand{\twothirds}{{\textstyle {2\over 3}}}
\newcommand{\half}{{\textstyle {1\over 2}}}
\newcommand{\fourth}{{\textstyle {1\over 4}}}
\newcommand{\eighth}{{\textstyle {1\over 8}}}
\newcommand{\twelfth}{{\textstyle {1\over 12}}}

\newcommand{\hyper}{\ {}_2 F_1
({ -\third,\, -\third;\, -\twothirds; \, {R_H^6 / \Rt^6}}) }
\newcommand{\hyperU}{\ {}_2 F_1  
({ -\third,\, -\third, -\twothirds; \, {U_H^3/ U^3}}) }
\newcommand{\gym}{g_{\rm YM}}
\newcommand{\QQ}{Q}
\newcommand{\Rt}{{R}}
\newcommand{\Bt}{{B}}
\newcommand{\bt}{{\beta}}
\newcommand{\Hh}{\hat{H}}
\newcommand{\R}{\mathrm{I}\kern -2.5pt \mathrm{R}}
\newcommand{\Z}{\mathsf{Z}\kern -4pt \mathsf{Z}}
\newcommand{\D}{{\rm d}}
\newcommand{\cO}{{\cal O}}
\newcommand{\SU}{\mathrm{SU}}
\def\th{\theta} 
\def\pa{\partial}
\def\e{ {\rm e} }
\def\al{\alpha}
\def\b{\beta}
\def\be{\beta}
\def\ga{\gamma} 
\def\de{\delta} 
\def\O{\Omega}
\def\L{\Lambda}
\def\theequation{\thesection.\arabic{equation}}

\begin{document}

\begin{flushright} 
{\tt hep-th/0412015}\\ 
BRX-TH-551\\ 
BOW-PH-133 \\
\end{flushright}
\vspace{30mm}

\begin{center}
{\bf\Large\sf Thermodynamics of the localized D2--D6 System}

\vskip 5mm
Marta~G\'omez-Reino\footnote{Research 
supported by the DOE under grant DE--FG02--92ER40706.}$^{,a}$,
Stephen G. Naculich\footnote{Research
supported in part by the NSF under grant PHY-0140281.}$^{,b}$,
and Howard J. Schnitzer\footnote{Research
supported in part by the DOE under grant DE--FG02--92ER40706.\\
{\tt \phantom{aaa} marta,schnitzer@brandeis.edu; naculich@bowdoin.edu}\\
}$^{,a}$

\end{center}

\vskip 1mm

\begin{center}
$^{a}${\em Martin Fisher School of Physics\\
Brandeis University, Waltham, MA 02454}

\vspace{.2in}

$^{b}${\em Department of Physics\\
Bowdoin College, Brunswick, ME 04011}
\end{center}

\vskip 2mm

\begin{abstract} 
An exact fully-localized extremal supergravity solution
for $N_2$ D2 branes and $N_6$ D6 branes,
which is dual to 3-dimensional supersymmetric $\SU(N_2)$
gauge theory with $N_6$ fundamentals,
was found by Cherkis and Hashimoto.
In order to consider the thermal properties of the gauge theory we 
present the non-extremal extension of this solution to first order in 
an expansion near the core of the D6 branes.
We compute the Hawking temperature and the black brane
horizon area/entropy. The leading order entropy, which is 
proportional to $N_2^{3/2} N_6^{1/2} T_H^2$,
is not corrected to first order in the expansion. 
This result is consistent with the analogous weak-coupling
result at the correspondence point $N_2 \sim N_6$.
\end{abstract}

\vfil\break

\section{Introduction}
\label{sec:intro}
\setcounter{equation}{0}
\renewcommand{\theequation}{1.\arabic{equation}}

An important issue for the gauge/gravity correspondence is how to
interpolate between strong gauge coupling (where the supergravity
approximation is valid) and the weakly-coupled regime of the gauge
theory.  In this context there have been several different
approaches to the search for string duals to large $N$ gauge
theories at weak coupling.  The study of the thermodynamics of
various systems is one tool in analyzing this question \cite{gkp,ft}. 
Typically
the thermal behavior of large $N$ gauge theories exhibits a
deconfining transition at sufficiently high temperature, which is
dual to a phase-transition from a thermal state to a black hole in
the string theory \cite{witten}. 
Such transitions can occur at both weak and
strong coupling, so that study of the thermal behavior of both
gauge theories and their dual string theories is well suited for
investigation of aspects of the gauge/gravity duality. 

One approach is to consider large $N$ gauge theories on 
a compact space, as this provides an additional parameter $R \L$, 
which may be tuned to weak
coupling, where $R$ is the size of the compact manifold, and $\L$
the dynamical scale of the gauge theory. References \cite{bo} and 
\cite{aharony} have provided a general framework for this
discussion on the gauge side. In that context, one can consider
$\SU(N)$ gauge theories with $N_f$ matter multiplets in the
fundamental representation of the gauge group, with $N_f/N$ finite
in the large $N$ limit.  It has been shown that in the
weak-coupling limit, this class of gauge theories on $S^{d-1}
\times$ {\it time} has two phases \cite{howard,sk}, separated by a third-order
phase transition at temperature $T_c$. The free energy in the
low-temperature phase behaves as
\beq
\frac{F}{T} \sim N^2_f \, f_{\rm low} (T)\, , \hspace{.5in} T \leq T_c\,,
 \eeq
whereas in the high-temperature phase, 
\beq
\label{eq:hitemp}
\frac{F}{T}  \sim N^2 \, f_{\rm high} \left( {N_f}/{N}\, , \;T \right) \, ,
\hspace{.3in} T \geq T_c\,,
 \eeq
where 
$f_{\rm low}  (T_c) = (N/N_f)^2 f_{\rm high}  (N_f/N \, , \; T_c)$. 
The high-temperature limit of (\ref{eq:hitemp}) becomes \cite{howard}
\beq\label{ht2}
F \sim  N^2 T^d \tilde{f} (N_f / N)\,.
\eeq
This can be interpreted as the behavior of 
glueballs and (color-singlet) mesons at low temperature,
with a deconfining transition to a phase of gluons and fundamental
(and anti-fundamental) matter at high temperature.  
It was
speculated \cite{howard} that the low-temperature phase was dual to a thermal
string state, with a high-temperature transition to a black hole.

In this paper, we study the thermal behavior of a string theory
dual to a gauge theory of this type.  
One might be tempted to consider the large $N$ limit of type IIB theory 
with $N$ D3 branes and $N_f$ D7 branes, with $N_f/N$ finite, 
but the conical singularities associated with D7 branes limits their number. 
A cleaner example, which can be used to study $d$=3 gauge theory, 
involves type IIA theory with $N_2$ D2 branes and $N_6$ D6 branes, 
where there is no restriction on the number of D6 branes.  
A fully localized D2--D6 solution is required for the dual 
to the $\SU(N_2)$ gauge theory with $N_6$ fundamentals.
In ref. \cite{cherkis} the
exact extremal ({\it i.e.}, zero-temperature) metric for
localized D2 and D6 branes was obtained.
Previously, the approximate (extremal) metric for a 
localized D2--D6 system valid near the core of the D6 branes 
was given in \cite{itzhaki}. This metric was also considered in 
\cite{ek} in the context of describing a gravity dual of mesons for this 
gauge theory.

In order to discuss the thermal properties of the gauge/string duality, 
one needs the non-extremal analog of the extremal solution presented 
in \cite{cherkis}. 
We have not been able to obtain an exact non-extremal solution 
of the localized D2-D6 system,  
so we consider a systematic expansion near the core of the D6 branes.
The non-extremal metric in the near-core region 
(corresponding to the IR fixed-point of the gauge theory) 
was obtained in \cite{pelc},
and corresponds to the leading term in our expansion.
In section 3, we obtain the first-order correction
to this non-extremal metric.
The correction to the metric involves the
beginning of the flow away from the IR fixed point.  

In section 4, we examine the thermodynamics implied 
by our solution in the decoupling limit, 
which uncouples the $\SU(N_2)$ gauge theory from the bulk, 
and leaves a $\SU(N_6)$ global flavor symmetry from the D6 branes. 
We compute the Hawking temperature $T_H$ and the 
area of the black-brane horizon,
which is proportional to the entropy,
for our solution.
We find the entropy as a function of temperature
\beq
\label{eq:main}
S =  \frac{8  \pi^2}{27} \sqrt{ 2N_2^3  N_6} \, V_2 \, T_H^2\,,
\eeq
valid to first-order in our expansion.
As is evident from (\ref{eq:main}), 
our calculation is that of
the high-temperature limit, {\it i.e.}, of the black-hole
thermodynamics in $d$=3.

The validity of the geometrical description requires
large $N_2 N_6$.
As one varies $N_6$ relative to $N_2$, the theory
has different phases \cite{pelc}.  
For small $N_6$, the theory is 11-dimensional, 
with geometry $AdS_4 \times S_7/\Z_{N_6}$.  
As $N_6/N_2$ increases, the geometry
becomes 10-dimensional, with $AdS_4$ fibered over a compact $X_6$.
When $N_6 \gg N_2$, the 10-dimensional geometry becomes highly
curved, and one passes to a weakly-coupled phase of the gauge
theory.  This suggests a ``correspondence point" between the
10-dimensional sugra regime and perturbative gauge theory \cite{howard},
estimated to be at
\beq
N_2 \sim N_6 \equiv N \; .
 \eeq
Applying this correspondence to (\ref{eq:main}), one has
\beq
S \sim N^2 \, T^2_H\,,
 \eeq
appropriate to a black-hole, and in agreement with the
high-temperature limit (\ref{ht2}).

\section{Extremal 11-dimensional supergravity solution}
\renewcommand{\theequation}{2.\arabic{equation}}
\setcounter{equation}{0}

A localized D2--D6 brane configuration in type IIA string theory 
uplifts to an M-theory configuration consisting
of M2 branes in a Taub--NUT background.
The supergravity solution corresponding to this configuration
satisfies the 11-dimensional equations of motion
\bea
\label{eq:ein}
R_{\mu\nu}  - \half g_{\mu \nu} R  
& = & 
\twelfth \left(  F_{\mu \al \be \ga} F_\nu{}^{\al \be \ga}
- \eighth g_{\mu \nu} F_{\al \be \ga \de} F^{\al \be \ga \de} \right)\,,
\\
\label{eq:Feom}
\D \, {}^*F_{[4]}  + \half  F_{[4]} \wedge F_{[4]}
&=& 0\,,
\eea
which follow from the bosonic part of the 11-dimensional supergravity action
\beq
\label{eq:action}
I = {1 \over 16 \pi G_{11} } 
\int
\D^{11} x \left[ \sqrt{ -g }
\left( R - \textstyle{{1\over 48}} F^2_{[4]} \right)
+ \textstyle{{1\over 6}} F_{[4]} \wedge F_{[4]}\wedge A_{[3]} \right]\,.
\eeq

In the absence of M2--branes, $F_{[4]}$ vanishes, 
and the flat space equations $R_{\mu\nu}  - \half g_{\mu \nu} R  = 0$
have as a solution the ($\R^7 \times $ Taub-NUT) metric, 
where the Taub-NUT metric may be written as
\beq
\D s^2_{TN}  =  
\left( 1 + \frac{2mN_6}{r} \right)
 \Big[ \, \D r^2 + r^2 (\D  \th^2 + \sin^2 \th \, \D \phi^2) \Big] 
  +  \frac{(4m)^2}{(1+\frac{2mN_6}{r})} \;
  \left( \D \psi + \frac{N_6}{2} \; \cos \th \,\D \phi \right)^2\,,
\eeq
with 
$0 \le \th <  \pi$, $0 \le \phi < 2 \pi$, and $0 \le \psi < 2 \pi$. 
The radius of the circle of the Taub-NUT metric at $r=\infty$ is 
$R_\# = 4 m$. 
This solution corresponds to the M-theory lift of a configuration
of $N_6$ coincident D6 branes located at $r=0$ and spanning $\R^7$.

If M2 branes are present, they act as a source for $F_{[4]}$.
The extremal supergravity solution for M2 branes 
in the ($\R^7 \times $ Taub-NUT) background was derived by Cherkis and 
Hashimoto in ref.~\cite{cherkis}. For $N_2$ parallel M2-branes 
spanning the $t$, $x^1$, $x^2$ directions they used the following ansatz
\bea
\label{eq:CHmetric}
\D s^2_{11} &=& H^{-2/3} (-\D t^2 + \D x^2_1 + \D x^2_2)
+ H^{1/3} (\D y^2 + y^2 \D \O^2_3 + \D s^2_{TN}) \,, \\[.1in]
\label{eq:Fansatz}
F_{[4]} &=& \D t \, \wedge \, \D x_1 \, \wedge \, \D x_2 \, \wedge \, 
\D H^{-1}\,,
\eea
where
\beq
\D \O_3^2  =  \D \alpha_1^2 + \sin^2 \alpha_1 \left( \D \alpha_2^2 +
 \sin^2 \alpha_2\, \D \alpha_3^2 \right)\,.
\eeq
If the M2 branes are coincident, 
and lie at $r=y=0$, 
the function $H$ only depends on $r$ and $y$.
Then the ansatz (\ref{eq:Fansatz}) substituted into eq.~(\ref{eq:Feom})
yields
\beq
\label{eq:Hry}
0 = \frac{1}{\sqrt{-g}} \pa_\mu \left( \sqrt{-g} F^{012\mu} \right)
= \frac{1}{ \left( 1+\frac{2mN_6}{r} \right) }
\frac{1}{r^2} \frac{\pa}{\pa r} \left( r^2 \frac{\pa}{\pa r} H(r,y) \right) + 
\frac{1}{y^3} \frac{\pa}{\pa y} \left( y^3 \frac{\pa}{\pa y} H(r,y) \right)\,. 
\eeq
In ref.~\cite{cherkis} this equation is explicitly solved  
as the Fourier transform of the confluent hypergeometric function. 

The aim of this paper is to find a non-extremal generalization
of this localized D2--D6 brane solution. 
In order to do so, the first step is to use the change of variables 
$r = {z^2}/{8m N_6}$ to rewrite 
the metric (\ref{eq:CHmetric}) as
\bea
\label{eq:yzmetric}
 \D s^2_{11} & = & H^{-2/3} (-\D t^2 + \D x^2_1 + \D x^2_2 )
 +  H^{1/3} \Bigg\{  
   \D y^2 
+  y^2 \D \O^2_3 
+  \left[ 1 + \left(\frac{z}{4mN_6} \right)^2  \right]  \D z^2 
\nonumber\\[.1in]
& + &  
 \left( \frac{z}{2} \right)^2 \; 
\left[ 1 + \left(\frac{z}{4mN_6} \right)^2  \right] \;
(\D \th^2 + \sin^2 \th \; \D \phi^2 )
\nonumber \\[.1in]
& + &  z^2 \left[ 1 +  \left(\frac{z}{4mN_6} \right)^2  \right]^{-1} \;
 \left(\frac{\D \psi}{N_6} + \textstyle{\frac{1}{2}} \: 
\cos \th \; \D \phi \right)^2 
 \Bigg\}\,.
\eea
In these variables the equation (\ref{eq:Hry}) for $H$ takes the form 
\beq
\label{eq:Hzy}
  \frac{1}{z^3} \frac{\pa}{\pa z} \left( z^3 \frac{\pa}{\pa z} H(z,y) \right) 
+ \frac{1}{y^3} \frac{\pa}{\pa y} \left( y^3 \frac{\pa}{\pa y} H(z,y) \right) 
= - \left( \frac{z}{4mN_6} \right)^2 
\frac{1}{y^3}\frac{\pa}{\pa y} \left( y^3 \frac{\pa}{\pa y} H(z,y)\right)\,. 
\eeq
One can solve this equation for $H(z,y)$ 
order-by-order in an expansion in $1/m$, 
with the leading order given by the solution found in 
refs.~\cite{itzhaki,pelc} for a localized D2-D6 system 
near the core of the D6 branes.

There is a further change of variables that can be made 
to simplify the computations. 
We seek variables $(\Rt, \bt)$ such that
\beq
\label{eq:df}
\D y^2 
+  \left[ 1 + \left(\frac{z}{4mN_6} \right)^2  \right]  \D z^2 
  = \D \Rt^2 + \Rt^2 \,  \D\bt^2\,.
 \eeq
The change of variables between the set $(y,z)$ and the set $(\Rt, \bt)$ 
is given by the relations
\beq
y  =  \Rt \cos \bt, \qquad f(z)  =  \Rt \sin \bt\,,
\eeq
where the function $f(z)$ is the solution of the differential equation
\beq
\D f = \sqrt{ 1 + \left(\frac{z}{4mN_6} \right)^2}  \,\, \D z \,,
\eeq
given by
\bea
\label{eq:fsolution}
f(z) &=& {z \over 2} \sqrt{ 1 + \left(\frac{z}{4mN_6} \right)^2 }
+ 2 m N_6 \, {\rm arcsinh} \left( z \over  4 m N_6 \right)
\nonumber\\[.1in]
 &=&  z 
\left[ 1 
       + \frac{1}{6} \left( z\over 4mN_6\right)^2 
       - {\frac{1}{40} } \left({z}\over{4mN_6}\right)^4  + \ldots
\right]\,.
\eea
Equation (\ref{eq:fsolution}) may be 
inverted in an expansion in $1/m$ to give
\beq
\label{eq:zf}
z =  \Rt \sin \bt \left[ 
1 - \frac{\Bt^2}{6m^2} 
  + \frac{13\Bt^4}{120m^4}  + \ldots \right] , \qquad
\Bt =  \frac{\Rt \sin\bt }{ 4 N_6}\,.
\eeq
Now, using  (\ref{eq:df}) and (\ref{eq:zf}),
one may rewrite the extremal metric (\ref{eq:yzmetric})
in an expansion in $1/m$
\bea
\label{eq:newmetric}
 \D s^2_{11} & = & H^{-2/3} (-\D t^2 + \D x^2_1 + \D x^2_2 )
 +  H^{1/3} \Bigg[  
    \D \Rt^2 
+ \Rt^2  \D \bt^2 
 +  (\Rt \cos\bt )^2  \D \O^2_3  
\nonumber\\
& + &  
\left( \frac{\Rt\sin\bt}{2} \right)^2 
\left(1  + \frac{2 \Bt^2}{3 m^2} - \frac{19 \Bt^4}{45 m^4} \right)
\; (\D \th^2 + \sin^2 \th \; \D \phi^2 )
\nonumber \\[.1in]
& + & (\Rt\sin \bt)^2
\left(1  - \frac{4 \Bt^2}{3 m^2} + \frac{86 \Bt^4}{45 m^4} \right)
 \left(\frac{\D \psi}{N_6} + \textstyle{\frac{1}{2}} \: 
\cos \th \; \D \phi \right)^2 
 \Bigg] 
+ {\cal O} \left(\frac{1}{m^6} \right)\,.
\eea
Higher order terms can be generated easily if needed.
In the limit $m \to\infty$,
the Taub-NUT space reduces to the orbifold $\R^4 / \Z_{N_6}$,
and the metric (\ref{eq:newmetric}) 
reduces to that obtained in \cite{itzhaki,pelc}.

The solution for $H$ may be written in 
an expansion in $1/m$ in the $(\Rt,\bt)$ variables:
\beq
H(\Rt,\bt ) = 1 + \QQ \sum^\infty_{n=0} \frac{h_n (\b
)}{R^{6-2n}(4mN_6)^{2n}}\, .
\eeq
Up to order $1/m^4$, we find
\beq 
\label{eq:Hnew}
H =  1 + \QQ \left[ \frac{1}{\Rt^6} + \frac{1}{15(4mN_6)^4 \Rt^2}
\left( 1 + \sin^2 \bt + \sin^4 \bt \right) \right]  
+ {\cal O} \left(\frac{1}{m^6} \right)\,,
\eeq
where $\QQ = 32\pi^2 \, N_2 N_6 \, \ell^6_p$ \cite{itzhaki}. 
Note that the 
first correction to $H$ occurs at $\cO(1/m^4)$ 
rather than $\cO(1/m^2)$. This will be useful for finding 
the non-extremal version of the metric (\ref{eq:newmetric})
in the next section.

\section{Non-extremal 11-dimensional supergravity solution}
\renewcommand{\theequation}{3.\arabic{equation}}
\setcounter{equation}{0}

We have not been able to find an exact non-extremal solution 
generalizing the extremal solution of ref. \cite{cherkis}, 
so we turn instead to find an approximate non-extremal solution, 
based on the $1/m$ expansion developed in the last section.  
We make the following ansatz for the non-extremal 
metric and antisymmetric tensor field
 \bea
\label{eq:nonextremalmetric}
 \D s^2_{11} & = & H^{-2/3} (-f_1 \D t^2 + \D x^2_1 + \D x^2_2 )
 +  H^{1/3} \Bigg[  
   f_1^{-1} \D \Rt^2 
+ \Rt^2  \D \bt^2 
 +  (\Rt \cos\bt )^2  \D \O^2_3  
\nonumber\\
& + &  
f_2 \left( \frac{\Rt\sin\bt}{2} \right)^2 
\left(1  + \frac{2\Bt^2}{3 m^2} \right)
\; (\D \th^2 + \sin^2 \th \; \D \phi^2 )
\nonumber \\[.1in]
& + & 
f_2^{-2}  (\Rt\sin \bt)^2
\left(1  - \frac{4\Bt^2}{3m^2} \right)
 \left(\frac{\D \psi}{N_6} + \textstyle{\frac{1}{2}} \: 
\cos \th \; \D \phi \right)^2 
 \Bigg] 
+ {\cal O} \left(\frac{1}{m^4} \right)\,,\\
F_{[4]} 
&=& \D t \, \wedge \, \D x_1 \, \wedge \, \D x_2 \, \wedge \, \D \Hh^{-1}\,.
\eea
Note that the function $\Hh$ in 
$F_{[4]}$ is distinct from the function $H$ in the metric.

First we consider the $m \to \infty$ limit,
where the ($\R^7 \times $ Taub-NUT) background reduces to 
$\R^7$ times the $\Z_{N_6}$ orbifold of $\R^4$. 
In that case, the non-extremal solution 
is given by \cite{pelc,cvetic} 
\bea
\label{eq:f}
f_1 &=& 1-\left(\frac{R_H}{\Rt} \right)^6 \, , \nonumber\\
f_2 &=& 1 \, ,\nonumber\\
H   &=& 1 + {\QQ \over \Rt^6}  \, ,
\qquad\qquad\qquad\qquad\qquad 
\qquad\qquad\qquad 
{\rm for~~} m = \infty \, .\nonumber\\
\Hh &=&  
\left[ 1-{\QQ \sqrt{1 + R_H^6/\QQ} \over \QQ + \Rt^6 }\right]^{-1} \, ,
\eea

The Hawking temperature associated with this black brane metric
is 
\beq
\label{eq:Hawking}
T_H =   \frac{3}{ 2 \pi R_H \sqrt{H(R_H)} }\,,
\eeq
found in the usual way by requiring the absence of 
a conical singularity at the horizon $\Rt = R_H$
of the Euclidean continuation.

Let us now consider the non-extremal solution away from $m = \infty$.
As seen from eq.~(\ref{eq:Hnew}),
$H$ receives no corrections at $\cO(1/m^2)$.
Using this we make the ansatz that, at $\cO(1/m^2)$,
$f_1$ and $\Hh$ are also unchanged from (\ref{eq:f}),
but that $f_2$ takes the form
\beq
f_2 = 1 + \left( \frac{\Rt}{4 m N_6} \right)^2 
f_R (\Rt) f_\be (\bt) + {\cal O} \left(\frac{1}{m^4} \right) \, .
\eeq
The Einstein equations (\ref{eq:ein}) can be written as
\beq
\label{eq:newein}
R_{\mu\nu} + g_{\mu\nu} R 
=  \twelfth  F_{\mu \al \be \ga} F_\nu{}^{\al \be \ga}\,.
\eeq
Since, to the order we are working, $\Hh$ only depends on $\Rt$,
the r.~h.~s.~only receives contributions from $t$, $x^1$, $x^2$, and $\Rt$
components. 
One can check that the l.~h.~s.~of (\ref{eq:newein}) vanishes
(to ${\cal O}(1/m^2)$) 
for all other components except the 
$\theta\theta$, $\phi\phi$, $\phi\psi$, and $\psi\psi$
components. 
Requiring that these components also vanish 
implies that $f_\be (\bt) = \sin^2 \bt $ and that $f_R (\Rt)$ satisfies
\beq
   3\,{\Rt}^2\,(\Rt^6 - {R_H}^6) \,f_R''(\Rt)
  \, + \,   3 \Rt \left( 11\,{\Rt}^6 - 5\,{R_H}^6 \right) \,f_R'(\Rt)
 \, -\,  12\,{R_H}^6\,f_R(\Rt)  
 = 8 {R_H}^6  \, .
\eeq
The solution to this equation may be written in 
terms of the hypergeometric function:
\beq
f_R(\Rt) = {2 \over 3} \left[ \hyper -1 \right]
\eeq
so therefore 
\beq
f_2(\Rt,\bt) = 1 + {2 \over 3} \left( \frac{\Rt}{4 m N_6} \right)^2 
\left[ \hyper -1 \right] \sin^2\bt + 
{\cal O} \left(\frac{1}{m^4} \right) \, .
\eeq
With this form for $f_2(\Rt,\bt)$, all the Einstein equations,
as well as the antisymmetric tensor field equation,
are satisfied at order $1/m^2$. 
Hence we have obtained the approximate non-extremal solution
\bea
\label{eq:newnonextremal}
 \D s^2_{11} & = & H^{-2/3} (-f_1 \D t^2 + \D x^2_1 + \D x^2_2 )
 +  H^{1/3} \Bigg\{  
   f_1^{-1} \D \Rt^2 
+ \Rt^2  \D \bt^2 
 +  (\Rt \cos\bt )^2  \D \O^2_3  
\nonumber\\
& + &  
\left( \frac{\Rt\sin\bt}{2} \right)^2 
\left[1  + \frac{2\Bt^2}{3m^2} \hyper \right]
\; (\D \th^2 + \sin^2 \th \; \D \phi^2 )
\nonumber\\[.1in]
& + & 
(\Rt\sin \bt)^2
\left[1  - \frac{4\Bt^2}{3m^2} \hyper \right]
 \left(\frac{\D \psi}{N_6} + \textstyle{\frac{1}{2}} \: 
\cos \th \; \D \phi \right)^2 
 \Bigg\} 
+ {\cal O} \left(\frac{1}{m^4} \right)
\nonumber \\ [.1in]
F_{[4]} 
&=& \D t \, \wedge \, \D x_1 \, \wedge \, \D x_2 \, \wedge \, \D \Hh^{-1}\, 
\eea
with
\bea
\label{eq:msquare}
f_1  &=& 1-\left(\frac{R_H}{\Rt} \right)^6 + {\cal O} \left(\frac{1}{m^4} \right)
\nonumber\\
H   &=& 1 + {\QQ \over \Rt^6}  + {\cal O} \left(\frac{1}{m^4} \right)
\nonumber\\
\Hh  &=&  
\left[ 1-{\QQ \sqrt{1 + R_H^6/\QQ} \over \QQ + \Rt^6 }\right]^{-1}
+ {\cal O} \left(\frac{1}{m^4} \right) 
\eea
The horizon of the black brane 
approximate solution (\ref{eq:newnonextremal})
is given by 
$\Rt = R_H + \cO(1/m^4)$.

We observe that the approximate solution (\ref{eq:newnonextremal})
will be valid as long as the correction term is smaller than
the leading term, thus $\Bt/m  \ll 1$ or $\Rt \ll 4 m N_6/ \sin \bt$.
However, the hypergeometric function multiplying the $1/m^2$  correction
diverges logarithmically at $\Rt = R_H$, 
so we must also have $\Rt \gg R_H$.  
This is unfortunate, as we are particularly interested in
the behavior of the solution at the horizon.
On the other hand, the fact that some components of the 
metric seem to diverge at the horizon may be an artifact 
of our $1/m$ expansion,
and it is conceivable that $f_2$, summed to all orders in $1/m$,
is perfectly regular.

Formally, the computation of the Hawking temperature 
of the metric (\ref{eq:nonextremalmetric})
is independent of $f_2$, 
and is given by eq.~(\ref{eq:Hawking}) through order $1/m^2$.
Thus, while our approximate metric clearly breaks down at the horizon,
there is no evidence of this breakdown in the Hawking temperature.

Furthermore, the area of the horizon of the black brane
computed using (\ref{eq:nonextremalmetric}),
\beq
\label{eq:horizonarea}
A =  \frac{ \pi^4 R_H^7}{3 N_6} \sqrt{ H(R_H) } V_2 \, ,
\qquad\qquad V_{2} = \int \D x_1 \D x_2
\eeq
is also formally independent of $f_2$,
and thus naively unaffected by the (apparent) divergence of $f_2$ 
at the horizon.   
By the Bekenstein-Hawking relation, the entropy $S = A/4 G_{11}$ 
is similarly  unaffected at order $1/m^2$. Therefore, 
while our $1/m^2$ approximation shows large corrections
to the metric near the black brane horizon 
(and in fact breaks down there),
both the Hawking temperature and the black brane horizon area are,
superficially at least, insensitive to the $1/m^2$ corrections 
and therefore given by their $m \to \infty$ values, without correction.

Higher orders in $1/m$ probably require a further generalization 
(beyond inclusion of the function $f_2$)
of the non-extremal metric and antisymmetric tensor field ansatz 
which may (or may not)
affect the value of the Hawking temperature and horizon area.
However, to $1/m^2$ at least, we see no evidence of any correction
to these quantities.
One could further speculate that the $m \to \infty$ results 
for the Hawking temperature and the horizon area/entropy 
are valid for all $m$.

\section{Decoupling limit}
\renewcommand{\theequation}{4.\arabic{equation}}
\setcounter{equation}{0}

Next, we consider the decoupling limit \cite{maldacena}
of our 11-dimensional approximate solution (\ref{eq:newnonextremal}).
For that purpose, we define 
\beq
\Rt^2 = \ell_p^3 U\,, \qquad
R_H^2 = \ell_p^3 U_H\,,
\eeq
where $\ell_p$ is the 11-dimensional Planck length.
In the decoupling limit $\ell_p \to 0$, with $U$ fixed, one has 
\beq
H= 1 + \frac{32\pi^2N_2N_6}{ \ell_p^3 U^3} \longrightarrow 
\frac{32\pi^2N_2N_6}{ \ell_p^3 U^3},
\eeq
yielding the following decoupled metric 
\bea
&& \ell_p^{-2} \D s^2_{11}  =  
U^2 (32\pi^2N_2N_6)^{-2/3} \left(-f_1 \D t^2 + \D x^2_1 + \D x^2_2 \right)
+ (32\pi^2N_2N_6)^{1/3}  \Bigg\{  f_1^{-1} \frac{ \D U^2 }{4 U^2} 
+ \D \bt^2 
\nonumber\\
&& +   \cos^2 \bt  \, \D \O^2_3  +
\fourth \sin^2 \bt
\left[1  + \frac{2 U \sin^2 \beta}{3 \gym^2 N_6^2}  \hyperU \right]
\; (\D \th^2 + \sin^2 \th \; \D \phi^2 )
\nonumber\\[.1in]
&& +  
\sin^2 \bt
\left[1  - \frac{4 U \sin^2 \beta}{3 \gym^2 N_6^2}  \hyperU \right]
 \left(\frac{\D \psi}{N_6} + \textstyle{\frac{1}{2}} \: 
\cos \th \; \D \phi \right)^2 
 \Bigg\}
+ {\cal O} \left(\frac{1}{m^4} \right)\,,
\nonumber\\
\label{eq:decouplingmetric}
\eea
where we have used $\ell_p^3 = g_s \ell_s^3,$ $4m = g_s \ell_s,$ and 
$\gym^2 = {g_s}/{\ell_s}$ 
(where $\gym$ is the 3$d$ gauge coupling 
on the D2-brane) to express
\beq
  \left( \frac{\Bt}{m} \right)^2  
= \left( \frac{\Rt \sin \bt}{4 m N_6} \right)^2
= \frac{U \sin^2 \bt}{\gym^2 N_6^2}  \, .
\eeq
{}From the decoupled metric (\ref{eq:decouplingmetric}) we can 
compute the Ricci scalar 
\beq
\label{eq:ricci}
R_{11}  =  \left( 27 \over 4 \pi^2 N_2 N_6  \right)^{1/3}  \frac{1}{\ell_p^2}
+ {\cal O} \left(\frac{1}{m^4} \right)\,.
\eeq
The validity of the 11-dimensional supergravity solution 
requires that the Ricci curvature $R_{11}$ 
be small compared to $\ell_p^{-2}$,
which is satisfied provided $N_2 N_6 \gg 1$. Moreover, the validity 
of the $1/m$ expansion employed in this 
paper requires that the corrections be small, i.e. $\Bt/m \ll 1$,
which implies
\beq
U \ll \frac{\gym^2 N_6^2}{\sin^2 \bt} \, .
\eeq
Simultaneously, we must impose $U > U_H$ 
since the $1/m$ expansion appears to break down near the
horizon $U \sim U_H$ due to a logarithmic divergence of 
the hypergeometric function.
(But note that the Ricci curvature (\ref{eq:ricci}) 
is unaffected by the $1/m^2$ corrections to the metric, 
and in particular is finite at the horizon.)

As we have noted above,
neither the Hawking temperature (\ref{eq:Hawking}) 
nor the horizon area (\ref{eq:horizonarea}) 
formally have any ${\cal O}(1/m^2)$ corrections.
If we assume that the divergence at $U=U_H$ is an artifact
of the approximation scheme, and that in fact
the Hawking temperature and horizon area are unchanged
to ${\cal O}(1/m^2)$,
we can compute their values in the decoupling limit to be
\beq
\label{eq:hawking}
T_H =   \frac{3 U_H}{ 8 \pi^2 \sqrt{ 2 N_2 N_6} }
\eeq
and\footnote{The $N_2$ and $N_6$ dependence of
our result differs from that in eq.~(3.23) in ref.~\cite{pelc}. }
\beq
A 
=  \frac{4 \pi^5}{3} \sqrt{ 2N_2 \over N_6}
\, V_2 \, \ell_p^9 \, U_H^2 
=  \frac{512 \pi^9}{27} \sqrt{ 2N_2^3  N_6}
\, V_2 \, \ell_p^9 \, T_H^2 \, .
\eeq
The horizon area is related to the entropy of the black brane
via the Bekenstein-Hawking relation
\beq
\label{eq:entropy}
S =  \frac{A}{4 G_{11} } 
=  \frac{8  \pi^2}{27} \sqrt{ 2N_2^3  N_6} \, V_2 \, T_H^2\,,
\eeq
where $G_{11} = 2^4 \pi^7 \ell_p^9$.
On the other hand, the breakdown of our approximation 
at the horizon may signal that higher-order effects 
are important, which may alter these conclusions.
We have not been able to compute the $1/m^4$ corrections to the
supergravity solution.

As a check on (\ref{eq:entropy}),
we may compute the entropy from the first law
of thermodynamics, following the method described in
refs.~\cite{thermo,witten}. 
The four-dimensional part of the decoupled metric (\ref{eq:decouplingmetric})
has the form of a black hole in $AdS_4$ with radius 
$\half \left( 32 \pi^2 N_2 N_6 \right)^{1/6} \ell_p $.
The Euclidean action for the black hole is given by
\beq
I = -\, {1 \over 16 \pi G_{4} } 
\int \D^{4} x \,  \sqrt{ -g_4 }  \left( R_4 - 2 \Lambda \right) \, ,
\eeq
where $R_4  = -48 / \left( 32 \pi^2 N_2 N_6 \right)^{1/3} \ell_p^2 $
and $\Lambda = - 12 / \left( 32 \pi^2 N_2 N_6 \right)^{1/3} \ell_p^2 $.
Evaluating the action using $1/G_4 = {\rm Vol}_7/G_{11} $, 
where $G_{11} = 2^4 \pi^7 \ell_p^9$
and 
${\rm Vol}_7 = (\pi^4/3N_6) \left( 32 \pi^2 N_2 N_6 \right)^{7/6} \ell_p^7$
we obtain 
\beq
I = 
\frac{V_2 } {64 \pi^4 N_6} 
\int_{U_H}^U  U^2 \, \D U 
\int_0^\beta \D \tau 
\,,
\eeq
where we have cut off the divergent integral at some large value $U$,
and where the period of the Euclidean time $\tau$ for 
the non-extremal metric is given by
\beq
 \beta = \frac{1}{T_H} = \frac{8\pi^2 \sqrt{2 N_2N_6} }{3U_H}\, .
\eeq
To obtain a finite result, one must subtract the 
action for the extremal metric
\beq
 I_{\rm e} = 
\frac{V_2 } {64 \pi^4 N_6} 
\int_0^U U^2 \: \D U
\int^{\beta'}_0 \D \tau  \, .
\eeq
Following refs.~\cite{thermo,witten}, 
one must adjust the period $\beta'$ of the extremal metric
so that the geometry is the same for extremal
and nonextremal metrics at the hypersurface at $U$.
This implies 
\beq
 \b' = \b \left[ 1- \left(\frac{U_H}{U}\right)^3 \right]^{1/2} \; .
\eeq
The regularized action is
\begin{eqnarray}
I - I_{\rm e}  = 
\frac{V_2 } {64 \pi^4 N_6} 
 \left[ \int^U_{U{_H}} U^2 \, \D U   \int^{\beta}_0 \D \tau 
- \int_0^U  U^2 \, \D U \int_0^{\beta'}  \D \tau \right] \; ,
 \end{eqnarray}
which, in the $U\rightarrow \infty$ limit, gives 
\beq
\Delta I = 
\lim_{U\rightarrow\infty} \left( I - I_{\rm e}\right)  = 
  \,-\, 
\frac{V_2 } {384 \pi^4 N_6} U_H^3   \beta
= \,-\, 
\frac{8 \pi^2 V_2 } {81} 
\sqrt{2 N_2^3 N_6} \beta^{-2}
\eeq
The energy is $ E = \frac{\pa }{\pa \b} \Delta I $, 
so that the entropy, as given by the first law of thermodynamics, is
 \beq
\label{entropy}
 S  =  \beta E -\Delta I 
 =  \frac{8  \pi^2}{27} \sqrt{ 2N_2^3  N_6} \, V_2 \, T_H^2\,,
\eeq
which agrees with the entropy (\ref{eq:entropy})
calculated using the Bekenstein-Hawking relation. This computation 
is valid through order $1/m^2$.

\section{Reduction to ten dimensions}
\renewcommand{\theequation}{5.\arabic{equation}}
\setcounter{equation}{0}

The $\psi$ direction is an isometry of the eleven-dimensional 
metric (\ref{eq:newnonextremal}), and so we can obtain the 
10-dimensional metric from the 11-dimensional one 
by dimensional reduction along the $\psi$ direction.\footnote{
In the absence of D6 branes, i.e., in a flat 11-dimensional background,
the M2 supergravity solution must be smeared in the $\psi$ direction
before it can be reduced.   In the presence of D6 branes, the
$\psi$ direction is ``pinched off'' at the D6 (and therefore D2) brane 
location, so no smearing is necessary.} The 10-dimensional 
metric in the string frame $\D s^2_{\rm str}$ and the dilaton $\phi$ are 
identified through the relation
 \beq
\label{eq:reduction}
\D s^2_{11}  = 
\e^{4 (\phi-\phi_\infty) /3} 
\left[ R_\# \, \D \psi + A_\mu \, \D x^\mu \right]^2
  +  \e^{-2(\phi-\phi_\infty)/3} \D s^2_{\rm str}
\eeq
where $R_\#$ is the asymptotic radius of the eleventh dimension
\beq
R_\#  =  4m \: = \: g_s\ell_s , \qquad g_s = \e^{\phi_\infty}\,.
\eeq
Comparing (\ref{eq:reduction}) with (\ref{eq:newnonextremal}), 
one obtains the dilaton
\beq
\label{eq:dilaton}
\e^{\phi} =  g_s  H^{1/4}  \left( \Bt \over m \right)^{3/2}  
\left[1  -  \frac{\Bt^2}{m^2} \hyper \right]
\eeq
and the 10-dimensional metric in the string frame
\bea
\label{eq:stringmetric}
\D s^2_{\rm str} &=&  
H^{-1/2} 
\frac{\Bt}{m} \left[1  -  {2\Bt^2 \over 3m^2} \hyper \right]
\left(-f_1 \D t^2 + \D x^2_1 + \D x^2_2 \right)
\nonumber\\
&+&  
H^{1/2}  \frac{\Bt}{m} \left[1  - {2\Bt^2 \over 3m^2}  \hyper \right]
\left( f_1^{-1} \D \Rt^2 + \Rt^2  \D \bt^2 
+  \Rt^2 \cos^2\bt  \D \O^2_3   \right)
\nonumber\\
& + &  
H^{1/2}  \frac{\Bt}{m} \left( \frac{\Rt\sin\bt}{2} \right)^2 
\; \left(\D \th^2 + \sin^2 \th \; \D \phi^2 \right)
+ {\cal O} \left(\frac{1}{m^4} \right) \, .
\eea
If we now take the decoupling limit, the dilaton becomes
\beq
\label{eq:decouplingdilaton}
\e^{\phi} =  \left( 32 \pi^2 N_2 \sin^6 \bt \over N_6^5 \right)^{1/4}  
\left[1  -  \frac{ U \sin^2 \beta}{ \gym^2 N_6^2}  \hyperU \right]
+ {\cal O} \left(\frac{1}{m^4} \right)
\eeq
and the 10-dimensional string metric becomes
\bea
\label{eq:decouplingstringmetric}
&&\D s^2_{\rm str} =  
\frac{\ell_s^2 \sin\bt}{N_6}
\Bigg\{ \frac{U^2}{(32\pi^2N_2N_6)^{1/2}}
\left[1  - \frac{2 U \sin^2 \beta}{3 \gym^2 N_6^2}  \hyperU \right]
\left(-f_1 \D t^2  + \right.
\nonumber\\
&& \left. \D x^2_1  + \D x^2_2 \right)+
 (32\pi^2N_2N_6)^{1/2}  
\left[1  - \frac{2 U \sin^2 \beta}{3 \gym^2 N_6^2}  \hyperU \right]
\left[  f_1^{-1} \frac{ \D U^2 }{4 U^2} 
+ \right.
\nonumber\\
&&\left. \D \bt^2 + \cos^2 \bt  \, \D \O^2_3  \right]+ 
\fourth (32\pi^2N_2N_6)^{1/2} \sin^2 \bt
\; \left(\D \th^2 + \sin^2 \th \; \D \phi^2 \right) \Bigg\}
+ {\cal O} \left(\frac{1}{m^4} \right) \, .
\eea
In the $m \to \infty$ limit,
the dilaton will be small provided
\beq
\frac{32 \pi^2 N_2 \sin^6 \bt}{N^5_6} \ll 1
\eeq
as discussed in ref.~\cite{pelc},
and the reduction to ten dimensions is only valid in that case.

Also, as discussed in the previous section, 
the $1/m$ expansion 
will be valid only if 
\beq
U_H \ll U \ll \frac{\gym^2 N_6^2}{\sin^2 \bt} \, .
\eeq
Since (unlike the Hawking temperature and horizon area)
the dilaton depends explicitly on $f_2$, 
whose $1/m^2$ contribution diverges (logarithmically) 
at the horizon, the reduction to 10-dimensional supergravity 
seems to be problematic, at least near the horizon,
although it seems legitimate away from the horizon.
(As mentioned above, $f_2$ summed to all orders in $1/m$ might
indeed be regular at the horizon, but we have no way
of knowing whether the dilaton remains small there.)

The ten-dimensional Ricci curvature computed from the
decoupled metric (\ref{eq:decouplingstringmetric}) is
\beq
\label{eq:ricciten}
R_{10}  =  {3\over 8 \pi} \sqrt{ N_6 \over 2 N_2 } 
\frac{(1 - 15 \cos 2 \beta)}{\sin^3 \beta}
\frac{1}{\ell_s^2}
+ {\cal O} \left(\frac{1}{m^2} \right)
\eeq
hence for $N_6 \gg N_2$, the ten-dimensional supergravity
solution is no longer applicable, and the theory is
described by a weakly-coupled field theory\cite{pelc}. 

An estimate of the transition 
from the supergravity description to that of the weakly-coupled gauge 
theory is $N_2 \sim N_6 = N$, 
which can be considered as a correspondence point. 
At this correspondence point, the black brane entropy 
(\ref{entropy}) goes as $S \sim N^2 T^2$,
consistent with the behavior expected for the 
high-temperature deconfining phase of the
gauge theory (\ref{ht2}).

Thus, we see that the non-extremal localized D2--D6 brane system
presents the expected high-temperature thermodynamics, which is
compatible with evolution from strong to weak 't~Hooft coupling,
{\it i.e.}, from small to large curvature in the geometry.    We 
also note that the calculation presented here gives just the
high-temperature limit of the system.  Therefore, we have no
opportunity to observe a possible Hawking--Page transition in the
bulk \cite{thermo,witten}, or the 3$^{\rm rd}$ order phase-transition 
found in the weakly-coupled gauge theory \cite{howard}.

\section{Discussion}
\renewcommand{\theequation}{6.\arabic{equation}}
\setcounter{equation}{0}

In this paper we have obtained a non-extremal version of 
the metric that describes the localized intersection 
of D2 and D6 branes given 
by Cherkis and Hashimoto in ref.~\cite{cherkis}.
The non-extremal version of this 
metric was found as a systematic expansion in the neighborhood of the 
core of the D6 branes (an expansion in $1/m$, where the parameter $m$ gives 
the radius of the Taub-NUT space), but to all orders in the non-extremal 
parameters. Certain restrictions were discussed which must be respected for 
the geometric description and $1/m$ expansion to be valid. It was found 
that this expansion breaks down near the horizon due to a logarithmic 
divergence of the hypergeometric function appearing in the 
non-extremal metric. 
Nevertheless, proceeding formally, we found that the Hawking temperature 
and Bekenstein-Hawking entropy 
do not depend on the hypergeometric function, so that 
in fact these thermodynamic quantities are uncorrected by terms of 
${\cal O} (1/m^2)$. One could argue thus that (\ref{eq:hawking}) and 
(\ref{eq:entropy}) will 
survive even with a better understanding of the non-extremal metric, 
as evidenced by the computation of the entropy from the action in 
(\ref{entropy}) and by the fact that (\ref{eq:entropy}) is compatible with 
the field theory result (\ref{ht2}) at the correspondence point 
$N_2 \sim N_6=N$. Also note that the non-extremal metric we have found 
corresponds to the high temperature limit, 
so that one is unable to consider 
a possible Hawking-Page transition in the bulk, which may be dual to 
the Gross-Witten phase transition \cite{gross} found in the weakly-coupled 
gauge theories \cite{howard}.

Of concern is that our systematic expansion breaks down near the horizon, 
as we already mentioned in the previous paragraph. We can envision at least 
two possibilities. The first is that the result is actually finite at the 
horizon when all orders in $1/m$ are included. The second is that this 
divergence is genuine, but may be cut-off by separating the D2 from the 
D6 branes, giving a small mass to the fields in the fundamental 
representation of the gauge group (the separation between D2 and D6 
branes was considered in \cite{cherkis} for the extremal case). One might then 
consider the logarithmic divergence (when the mass of the fundamentals 
vanishes) as a delocalization effect analogous to that discussed by 
\cite{marolf}. These are issues for future study.

Another possible extension of this work would be to study the interpolation 
between strong and weak coupling along the lines of ref.~\cite{gkp}, 
in which the free energy for the four-dimensional 
${\cal N}=4$ $\SU(N)$ gauge theory is studied. In the large $N$ limit, 
the entropy is $N^2T^3$ times a function of the 
't~Hooft coupling $f(\gym^2 N)$.  
In the strong-coupling limit, the entropy was found to be 3/4 that of the
weak-coupling limit, with corrections of order $(\gym^2 N)^{-3/2}$, 
coming from $R^4$ terms in the IIB effective string action.
In our case, we anticipate an analogous, but richer situation,
where we expect our strong-coupling result (\ref{eq:main}) 
to be corrected by a function 
$f(\gym^2 N_2/U, \; N_6/N_2)$, interpolating
between strong (IR) and weak (UV) domains.

\section*{Acknowledgments}

We would like to thank Stanley Deser, Martin Kruczenski,
Albion Lawrence, Hong Liu, Carlos Nu\~ nez,
Malcolm Perry, and Anton Ryzhov 
for helpful conversations.
MG-R and HJS wish to thank the 2nd Simons Workshop 
at Stony Brook for hospitality and a stimulating
atmosphere.
HJS wishes to thank the string group 
at the Harvard University Physics Department 
for hospitality extended over a long period of time.


\begin{thebibliography}{99}

\bibitem{gkp}
S.~S.~Gubser, I.~R.~Klebanov and A.~W.~Peet,
``Entropy and Temperature of Black 3-Branes,''
Phys.\ Rev.\ D {\bf 54}, 3915 (1996)
[arXiv:hep-th/9602135]. %%CITATION = HEP-TH 9602135;%%
I.~R.~Klebanov and A.~A.~Tseytlin,
``Entropy of Near-Extremal Black p-branes,''
Nucl.\ Phys.\ B {\bf 475}, 164 (1996)
[arXiv:hep-th/9604089].  %%CITATION = HEP-TH 9604089;%%
 S.~S.~Gubser, I.~R.~Klebanov and A.~A.~Tseytlin,
``Coupling constant dependence in the thermodynamics of N = 4  supersymmetric 
Yang-Mills theory,''
Nucl.\ Phys.\ B {\bf 534}, 202 (1998)
[arXiv:hep-th/9805156]. %%CITATION = HEP-TH 9805156;%%

\bibitem{ft}
A.~Fotopoulos and T.~R.~Taylor,
``Comment on two-loop free energy in N = 4 supersymmetric Yang-Mills  theory
 at finite temperature,''
Phys.\ Rev.\ D {\bf 59}, 061701 (1999)
[arXiv:hep-th/9811224]. %%CITATION = HEP-TH 9811224;%%

\bibitem{witten}
E.~Witten,
``Anti-de Sitter space, thermal phase transition, and confinement in  gauge
theories,''
Adv.\ Theor.\ Math.\ Phys.\  {\bf 2}, 505 (1998)
[arXiv:hep-th/9803131]. %%CITATION = HEP-TH 9803131;%%

\bibitem{bo}
B.~Sundborg,
``The Hagedorn transition, deconfinement and N = 4 SYM theory,''
Nucl.\ Phys.\ B {\bf 573}, 349 (2000)
[arXiv:hep-th/9908001]. %%CITATION = HEP-TH 9908001;%%

\bibitem{aharony}
O.~Aharony, J.~Marsano, S.~Minwalla, K.~Papadodimas and M.~Van Raamsdonk,
``The Hagedorn/deconfinement phase transition in weakly-coupled large N
gauge theories,''
arXiv:hep-th/0310285. %%CITATION = HEP-TH 0310285;%% 
H.~Liu,
``Fine structure of Hagedorn transitions,''
arXiv:hep-th/0408001. %%CITATION = HEP-TH 0408001;%%

\bibitem{howard}
H.~J.~Schnitzer,
``Confinement/deconfinement transition of large $N$ gauge theories with $N_f$
fundamentals: $N_f/N$ finite,''
Nucl.\ Phys.\ B {\bf 695}, 267 (2004)
[arXiv:hep-th/0402219]. %%CITATION = HEP-TH 0402219;%%

\bibitem{sk}
B.~S.~Skagerstam,
``On The Large $N_c$ Limit Of The SU$(N_c)$ Color Quark - Gluon Partition
Function,''
Z.\ Phys.\ C {\bf 24}, 97 (1984). %%CITATION = ZEPYA,C24,97;%%
A.~Dumitru, J.~Lenaghan and R.~D.~Pisarski,
``Deconfinement in matrix models about the Gross-Witten point,''
arXiv:hep-ph/0410294. %%CITATION = HEP-PH 0410294;%%

\bibitem{cherkis}
S.~A.~Cherkis and A.~Hashimoto,
``Supergravity solution of intersecting branes and AdS/CFT with flavor,''
JHEP {\bf 0211}, 036 (2002)
[arXiv:hep-th/0210105]. %%CITATION = HEP-TH 0210105;%%

\bibitem{itzhaki}
N.~Itzhaki, A.~A.~Tseytlin and S.~Yankielowicz,
``Supergravity solutions for branes localized within branes,''
Phys.\ Lett.\ B {\bf 432}, 298 (1998)
[arXiv:hep-th/9803103]. %%CITATION = HEP-TH 9803103;%%

\bibitem{ek}
J.~Erdmenger and I.~Kirsch,
``Mesons in gauge/gravity dual with large number of fundamental fields,''
arXiv:hep-th/0408113. %%CITATION = HEP-TH 0408113;%%

\bibitem{pelc}
O.~Pelc and R.~Siebelink,
``The D2-D6 system and a fibered AdS geometry,''
Nucl.\ Phys.\ B {\bf 558}, 127 (1999)
[arXiv:hep-th/9902045]. %%CITATION = HEP-TH 9902045;%%

\bibitem{cvetic}
M.~Cvetic and A.~A.~Tseytlin,
``Non-extreme black holes from non-extreme intersecting M-branes,''
Nucl.\ Phys.\ B {\bf 478}, 181 (1996)
[arXiv:hep-th/9606033].  %%CITATION = HEP-TH 9606033;%%

\bibitem{maldacena}
J.~M.~Maldacena,
``The large N limit of superconformal field theories and supergravity,''
Adv.\ Theor.\ Math.\ Phys.\  {\bf 2}, 231 (1998)
[Int.\ J.\ Theor.\ Phys.\  {\bf 38}, 1113 (1999)]
[arXiv:hep-th/9711200]. %%CITATION = HEP-TH 9711200;%%

\bibitem{thermo}
S.~W.~Hawking and D.~N.~Page,
``Thermodynamics Of Black Holes In Anti-De Sitter Space,''
Commun.\ Math.\ Phys.\  {\bf 87}, 577 (1983).  %%CITATION = CMPHA,87,577;%%

\bibitem{marolf}
D.~Marolf and A.~W.~Peet,
``Brane baldness vs. superselection sectors,''
Phys.\ Rev.\ D {\bf 60}, 105007 (1999)
[arXiv:hep-th/9903213]. %%CITATION = HEP-TH 9903213;%%

\bibitem{gross}
D.~J.~Gross and E.~Witten,
``Possible Third Order Phase Transition In The Large N Lattice Gauge Theory,''
Phys.\ Rev.\ D {\bf 21}, 446 (1980). %%CITATION = PHRVA,D21,446;%%

\end{thebibliography}
\end{document}